\documentclass[11pt,a4paper]{article}
\usepackage{amsmath,amssymb,mathrsfs}
\usepackage{graphicx}
\usepackage{comment, cite}

\begin{document}
\begin{titlepage}
\begin{center}
\leavevmode \\
\vspace{-2 cm}

{\scriptsize 
\hfill KEK-Cosmo-179\\
\hfill KEK-TH-1846\\
\hfill RIKEN-QHP-195\\
\hfill RIKEN-STAMP-11\\
\hfill UT-15-23 \\
\hfill DESY 15-114\\
\hfill COSMO-KOBE-16-13\\
\hfill  APCTP Pre2016-20\\
\hfill KIAS-P16075 \\
}

\noindent
\vskip 1 cm
{\LARGE \bf Reinterpretation of the Starobinsky model}

\vglue .5in

{\large
Takehiko Asaka$^1$, Satoshi Iso$^{2, 3}$, Hikaru Kawai$^4$, \\ Kazunori Kohri$^{2, 3}$, Toshifumi Noumi$^{5, 6, \dag}$, Takahiro Terada$^{7, 8, 9, \ddag}$
}

\vglue.3in

{\em \footnotesize
${}^1$ Department of Physics, Niigata University, Niigata 950-2181, Japan, \\
${}^2$ KEK Theory Center, High Energy Accelerator Research Organization (KEK), 
	Ibaraki 305-0801, Japan,\\
${}^3$ Graduate University for Advanced Studies (SOKENDAI), Ibaraki 305-0801, Japan, \\
${}^4$ Department of Physics, Kyoto University, Kyoto 606-8502, Japan, \\
${}^5$ Theoretical Research Division, Nishina Center, RIKEN, Wako 351-0198, Japan, \\
${}^6$ Jockey Club Institute for Advanced Study, Hong Kong University of Science and Technology, Hong Kong, \\
${}^7$ Department of Physics, The University of Tokyo, Tokyo 113-0033, Japan, \\
${}^8$  Deutsches Elektronen-Synchrotron (DESY), 22607 Hamburg, Germany, \\
${}^9$ Asia Pacific Center for Theoretical Physics (APCTP), Pohang 37673, Republic of Korea, \\
${}^{\dag}$ {\rm current address:} Department of Physics, Kobe University, Kobe 657-8501, Japan, \\
${}^{\ddag}$ {\rm current address:} Korea Institute for Advanced Study (KIAS), Seoul 02455, Republic of Korea \\
}

\end{center}

\vglue.4in

\begin{abstract}
The Starobinsky model of inflation, consistent with Planck 2015, has a peculiar form of the action, which contains the leading Einstein term $R$, the $R^2$ term with a huge coefficient, and negligible higher-order terms.
We propose an explanation of this form based on compactification of extra dimensions. 
Once tuning of order $10^{-4}$ is accepted to suppress the linear term $R$, we no longer have to suppress higher-order terms, which give nontrivial corrections to the Starobinsky model.
We show our predictions of the spectral index, its runnings, and the tensor-to-scalar ratio.
Finally, we discuss a possibility that quantum gravity may appear at the scale $\Lambda \gtrsim 5 \times 10^{15}$ GeV.
\end{abstract}
\end{titlepage}

\section{Introduction}
The precise cosmic microwave background (CMB) observations favor the plateau-type inflaton potentials.
Although the combined analysis of BICEP--Keck-Array--Planck resulted in a finite value of the tensor-to-scalar ratio, $r=0.048^{+0.035}_{-0.032}$~\cite{Ade:2015tva}, Planck 2015 itself has not found any evidence of detecting it but obtained an upper bound, $r<0.103$ (Planck TT $+$ lowP)~\cite{Ade:2015lrj}.
In fact, combining these results in a stringent limit, we get $r<0.08$ (Planck TT$+$lowP$+$BKP)~\cite{Ade:2015lrj}.
There are many models, including the Starobinsky model~\cite{Starobinsky:1980te}, the Higgs inflation model~\cite{Bezrukov:2007ep}, and cosmological attractors (see Ref.~\cite{Galante:2014ifa} and references therein),  whose predictions are at the center of the Planck constraint.
Among other features, the Starobinsky model of inflation specifically does not require introduction of an inflaton field by hand: the inflaton degree of freedom emerges from a higher-order gravitation term.

The well-known form of the model\footnote{
The original formulation involves all quadratic curvature invariants such as $R^{\mu \nu}R_{\mu \nu}$ and $R^{\mu\nu\rho\sigma}R_{\mu \nu \rho \sigma}$.  In conformally flat space-time, the effects of these terms are represented by the scalar curvature term as in Eq.~\eqref{R2}.
Note also that the de Sitter expansion in $f(R)$ gravity was discussed in Ref.~\cite{Barrow:1983rx}.
} is~\cite{Mijic:1986iv}
\begin{align}
S= \int d^4 x \sqrt{-g} \left(-\frac{1}{2} M_{\text{P}}^2 R + \frac{M_{\text{P}}^2}{12 m^2}R^2 \right) , \label{R2}
\end{align}
where $M_{\text{P}}\simeq 2.4 \times 10^{18}$ GeV is the reduced Planck mass, $R$ is the Ricci scalar, and $m$ is a mass-dimensional parameter (actually the inflaton mass).
In the IR limit, $R \ll m^2$, it reduces to the General Relativity (with the cosmological constant, which should be fine-tuned to be a small number and hence we ignore it here), which is well established in wide scales.
On the other hand, when $R$ becomes comparable with $m^2$, the second term becomes important.
Introducing an auxiliary scalar field and applying Weyl transformation and scalar field redefinition, the model is recast in the form of Einstein gravity with a canonically normalized scalar field $\phi$ with the following scalar potential~\cite{Whitt:1984pd, Maeda:1987xf, Barrow:1988xi}:
\begin{align}
V_{\text{Starobinsky}}=\frac{3}{4}m^2 M_{\text{P}}^2 \left(1-e^{-\sqrt{2/3}\phi /M_{\text{P}}}\right)^2, \label{R2_potential}
\end{align}
where $m$ is interpreted as the inflaton mass at the vacuum. 

If one interprets $m^2$ in Eq.~\eqref{R2} as the 
expansion parameter of the theory, there are no reasons to expect absence of even higher-order terms like $R^3$ and $R^4$ (aside from terms involving Ricci and Riemann tensors and derivatives, which we neglect because they generically introduce negative norm states (ghosts)~\cite{Stelle:1976gc}) with negative powers of $m^2$.
Such higher-order terms are extensively discussed in non-supersymmetric~\cite{Saidov:2010wx, Huang:2013hsb, Sebastiani:2013eqa, Codello:2014sua, Ben-Dayan:2014isa, Artymowski:2014gea, Rinaldi:2014gua, Broy:2014xwa, Artymowski:2015mva} as well as supergravity theories~\cite{Cecotti:1987sa, Farakos:2013cqa, Ferrara:2013kca, Ketov:2013dfa, Ozkan:2014cua, Diamandis:2015xra}.
That is, Eq.~\eqref{R2} should be augmented by higher-order terms as follows:
\begin{align}
S= M_{\text{P}}^2 \int d^4 x \sqrt{-g} \left( -\frac{1}{2}R+\sum_{n=2}^{\infty} a_{n} m^2 \left( \frac{R}{m^{2}} \right)^n  \right),\label{R2_higher}
\end{align}
where $a_2= 1/12$, and with $a_n$ ($n\geq 3$) naively expected to be of order $1$.
These terms, however, easily spoil the success of the inflationary model by substantially modifying the inflaton potential~\eqref{R2_potential}.
So in any way, 
the higher-order terms must be sufficiently suppressed to maintain the predictions of the model.

If the higher-order terms involving negative powers of $m^2$ are suppressed by phenomenological reasons, what is the scale of the suppression?
The ``next-to-natural'' expectation would be that it is the reduced Planck scale $M_{\text{P}}$, since there are no other scales in the theory.
In this case, the action is expanded by the Planck scale $M_{\text{P}}$ with order $1$ coefficients, but then the coefficient of the second term $R^2$ must be somewhat large ($a_2\simeq 5\times 10^8$).
This is the well-known peculiarity of the Starobinsky model, which we try to partially explain here.

In this paper, we take a view that the large coefficient of the $R^2$ term is actually an overall coefficient of the action. 
As we will see, such a large overall factor naturally emerges in theories with extra space-time dimensions.
(See Refs.~\cite{Gunther:2003zn, Bronnikov:2007qt, Saidov:2008tb, Apostolopoulos:2010jg, Briscese:2013lna} and references therein for previous works on extensions of the Starobinsky model in extra dimensions.)
Although we have to suppress the coefficient of the linear term $R$ by tuning of order $10^{-4}$, we do not have to additionally suppress the higher-order terms.  Moreover, it leads to a  Starobinsky-like model with interesting observational consequences.
At the end, we predict inflationary observables, and obtain the lower bound on the fundamental scale of the underlying higher-dimensional theory.

%%%%%%%%%%%%%%%%%%%%%%%%%%%%%%%%%%%%%%%%%
\section{Starobinsky-like model from extra dimensions}
Suppose that the underlying gravitational theory lives in $D$ space-time dimensions with a characteristic energy scale $\Lambda$.
Its effective action is described by
\begin{align}
S= \Lambda^{D} \int d^D x \sqrt{-g_D} \sum_{n=0} b_n  \left( \frac{R_D}{\Lambda^{2}} \right)^n, \label{DdimAction}
\end{align}
where $b_n$ are dimensionless coefficients, $g_D$ is the determinant of the $D$-dimensional metric, and $R_D$ is the $D$-dimensional Ricci scalar.
Here we require the absence of ghosts and higher derivative interactions.
We also require that the low-energy modes of our effective field theory are the graviton and the scalaron.
Hence we assume the $f(R)$-type theory (see \textit{e.g.} Refs.~\cite{ Sotiriou:2008rp, DeFelice:2010aj, Nojiri:2010wj} for reviews of $f(R)$ gravity).
Also, we neglect possible nonminimal couplings with matter fields for simplicity.
Assuming $b_2 >0$, we may set $b_2=1$ by redefinition of the scale $\Lambda$ or by Weyl transformation.
It should be stressed that this is not tuning but just a matter of convention.
(The $b_2$-dependence can be easily reproduced by replacing $b_1$ by $b_1/b_2$ after Eq.~\eqref{first_2parm_fit}.)
Upon the compactification to four dimensions, the action becomes
\begin{align}
S=c \int d^4 x \sqrt{-g} \sum_{n=0} b_n \Lambda^4  \left( \frac{R}{\Lambda^{2}} \right)^n, \label{CompactifiedAction}
\end{align}
where $c\equiv V_{D-4}\Lambda^{D-4}$ is the overall dimensionless factor, $V_{D-4}$ is the volume of the compactified extra dimensions, and $R$ is the four-dimensional Ricci scalar\footnote{
The above compactification assumes flat extra dimensions so that $R_D=R$.
For generic extra dimensions, we have $R_D= R + \mathcal{O}(1/L^2)$ where $L$ is the typical size of extra dimensions (compactification radius).  Thus, coefficients $b_n$ receive only corrections like $b_n \to b_n + \mathcal{O}\left( (1/L^2 \Lambda^2) \right)$.  Since we take $L\Lambda$ large, these corrections are neglected. 
}.
For example, if we take $D=10$ (\textit{cf.} superstring theory) and the compactification radius $L \equiv V_{6}^{1/6}$, which satisfies $L\simeq 30 / \Lambda$, we can naturally obtain a large overall factor $c \simeq 5\times 10^8$.
Note that if we take a larger number of dimensions, it becomes easier to obtain huge overall factor $c$ with the compactification radius being the same order as $\Lambda^{-1}$.

Basically, all the coefficients $b_n$ are expected to be order $1$, but $b_0$ should be fine-tuned to suppress the cosmological constant.
Furthermore, we require that $b_1$ also happens to be very small.
Otherwise, the situation is similar to Eq.~\eqref{R2_higher} with $m$ replaced by $\Lambda$.
Since the limit $b_1\to 0$ does not enhance symmetries of the underlying theory, it is regarded as tuning.
(However, in the perspective of the compactified four-dimensional theory, an approximate scale symmetry appears in this limit.  See more discussion in Sect.~\ref{sec:discussion}.)
The tuning is to be done so that the \emph{renormalized} value of $b_1$ becomes small. 
To reproduce the Starobinsky model, we set
\begin{align}
  c b_1 \Lambda^2 =&- \frac{M_{\text{P}}^2}{2}, & c=& \frac{M_{\text{P}}^2}{12m^2}\simeq 5 \times 10^8, \label{first_2parm_fit}
\end{align}
with $|b_1| \ll 1$.
Then Eq.~\eqref{CompactifiedAction} becomes
\begin{align}
S= \int d^4 x \sqrt{-g} \left(-\frac{1}{2}M_{\text{P}}^2 R + \frac{M_{\text{P}}^2}{12m^2}\left( R^2 + \sum_{n=3}^{\infty} b_{n} \left(- \frac{6 m^2}{b_1}\right)^{2-n} R^n \right) \right),  \label{newR2}
\end{align}
so the suppression scales for $n\geq 3$ become larger than the inflation scale $m$ as we tune $b_1$ to be small.
This is the form of the action we advocate.

The extension of the Starobinsky model with an $R^n$ term was studied in Ref.~\cite{Huang:2013hsb}, which gives us a constraint on each coefficient,
\begin{align}
\left| n b_n \left( \frac{b_1}{2} \right)^{n-2} \right| \lesssim  10^{-2n+2.6} \qquad (n\geq  3). \label{b1bn_bound}
\end{align}
The constraint for $n=3$ is
$|b_1| \lesssim 10^{-3.6}$ up to an order $1$ factor $|b_3|$.
With $|b_1|$ satisfying this bound, the constraint~\eqref{b1bn_bound} is also satisfied for $n\geq 4$.
Using the results of Planck 2015, $n_{\text{s}}=0.9655 \pm 0.0062$ (Planck TT $+$ low P)~\cite{Ade:2015lrj}, we obtain more stringent 95\% confidence-level bounds on $b_1$:
\begin{align}
-2.5 \times 10^{-4} \lesssim b \lesssim 1.3 \times 10^{-4} \qquad (N_e =50), \label{b_bound50} \\
-1.4 \times 10^{-4} \lesssim b \lesssim 2.2 \times 10^{-4} \qquad (N_e =60), \label{b_bound60}
\end{align}
where $b=b_3 b_1$. Note that $b_1<0$, and $b_3$ is expected to be of order $1$, so the bound on $b_1$ is roughly $|b_1|\lesssim 2 \times 10^{-4}$.\footnote{ \label{ftn:tuning}
One can soften the tuning of $b_1$ by an order of magnitude by tuning $b_3$ instead while keeping the value of $b$.  Further increase of $|b_1|$ makes the total tuning worse, which may be defined as the product of all $|b_n|$ less than one.  If we accept $|b_n|$ as small as the combinatorial factor $1/n!$, $|b_1|$ can be increased one more order of magnitude.
}

Let us compare how the situation has been improved.
The coefficients $a_n$ of the Starobinsky-like models in effective field theory~\eqref{R2_higher} should obey an constraint analogous to the inequality~\eqref{b1bn_bound} with $|b_1|$ replaced by an order one number.
The first higher-order coefficient $a_3$ is then constrained as severely as $b$, \textit{i.e.}, $|a_3|\lesssim \mathcal{O}(10^{-4})$.
However, the required tuning becomes increasingly more sever in the large $n$ limit, and an infinite tuning is required with $n\to\infty$.\footnote{
It may be possible for some higher-order terms to conspire to cancel major effects on the inflationary potential, but we do not discuss this case for simplicity.
}
In our proposal, only one parameter tuning of $b_1$ is needed, and its precision level is that of the least severe term in the original framework.

%%%%%%%%%%%%%%%%%%%%%%%%%%%%%%%%%%%%%%%%%
\section{Implications for inflationary observables}
For self-completeness and with the latest Planck data, let us discuss the effects of the extra terms in the action on the inflationary observables, namely the scalar spectral index $n_\text{s}$, its running $\alpha_{\text{s}}$, and the tensor-to-scalar ratio $r$.
As we saw above, effects of higher-order $(n\geq 4)$ terms are more suppressed than the $n=3$ term, so we neglect the higher-order terms in the following analyses.
The addition of the $R^3$ term has been studied since the early days~\cite{Barrow:1988xh, Berkin:1990nu}, but recent discussions include Refs.~\cite{Huang:2013hsb, Broy:2014xwa}.
Here, we summarize the properties of the scalar potential in the Einstein frame and inflationary observables, and compare with the latest observational data.
Considering observation of the 21 cm line from hydrogen atoms, we obtain the opposite conclusion to that in the literature~\cite{Huang:2013hsb}.

Under the standard procedure, the Jordan frame action~\eqref{newR2} up to the third term is transformed into the Einstein frame action with the following potential for a canonical scalar field $\phi$,
\begin{align}
V=\frac{m^2}{9 b^2}e^{-2\sqrt{2/3}\phi} \left( \sqrt{1+3b \left( e^{\sqrt{2/3}\phi}-1 \right)} -1 \right) \left( 1+6b \left( e^{\sqrt{2/3}\phi}-1 \right) -\sqrt{ 1+3b \left( e^{\sqrt{2/3}\phi}-1 \right)}  \right). \label{V}
\end{align}
Here and hereafter, we take the reduced Planck unit $M_{\text{P}}=2.4\times 10^{18} \text{GeV}=1$.
When we set $b=0$, it reduces to the Starobinsky potential~\eqref{R2_potential}.
If $b$ is negative, the potential blows up in the large field region.
If $b$ is positive, the potential has a runaway behavior, and there is the possibility of topological inflation as discussed in Ref.~\cite{Kamada:2014gma}.
If we retain $b$ up to the leading nontrivial order, the potential is
\begin{align}
V=V_{\text{Starobinsky}}\times \left( 1-\frac{b}{2}e^{\sqrt{2/3}\phi}\left(1-e^{-\sqrt{2/3}\phi} \right) \right) + \mathcal{O}(b^2),
\end{align}
where $V_{\text{Starobinsky}}$ is given in Eq.~\eqref{R2_potential}.
Of course, the large field behavior $(\phi \rightarrow \infty)$ depends also on the higher-order terms~\cite{Sebastiani:2013eqa, Broy:2014xwa}, but the leading order is enough for our purpose.

In the leading order of the deformation parameter $b$, 
 the spectral index $n_{\text{s}}$, the tensor-to-scalar ratio $r$, the running of the spectral index $\alpha_{\text{s}}$, and its running $\beta_{\text{s}}$ are obtained as
\begin{align}
1-n_{\text{s}} \simeq & \frac{2}{N} \left( 1 + \frac{16}{27}b N^2 \right) = \frac{2}{N}+\frac{32}{27}bN, \label{ns}\\
r \simeq & \frac{12}{N^2}\left( 1 -\frac{16}{27}b N^2 \right)=\frac{12}{N^2} - \frac{64}{9}b, \label{r}\\
\alpha_{\text{s}} \simeq & -\frac{2}{N^2} \left( 1-\frac{16}{27}bN^2 \right) = -\frac{2}{N^2}+\frac{32}{27}b, \label{alphas} \\
\beta_{\text{s}} \simeq & -\frac{4}{N^3} \left( 1+\frac{4}{9}bN \right) = - \frac{4}{N^3}-\frac{16}{9N^2}b, \label{betas}
\end{align}
where higher-order terms in $1/N$ are also neglected.
The results for $n_{\text{s}}, r$, and $\alpha_{\text{s}}$ are consistent with the $n=3$ case in Ref.~\cite{Huang:2013hsb}, and we additionally obtain the expression for $\beta_{\text{s}}$.

\begin{figure}[tbhp]
  \begin{center}
    \includegraphics[clip,width=10.0cm]{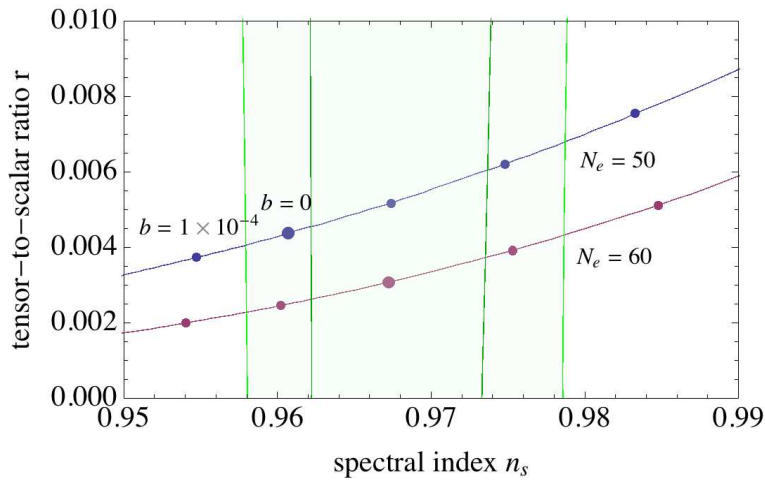}
    \caption{Prediction of the model in the ($n_{\text{s}}, r$)-plane as we vary the parameter $b=b_3 b_1$. 
    The blue (top) line and red (bottom) line correspond to $N_{e}=50$ and $N_{e}=60$, respectively.
    The large dots show the prediction of the Starobinsky model ($b=0$), and the small dots correspond to points of $b= n \times 10^{-4}$ with integer $n$.
    The green contours are the Planck TT+lowP+BKP+lensing+BAO+JLA+$H_0$ constraints (traced from Fig. 21 in Ref.~\cite{Planck:2015xua}).}
    \label{fig:ns_r}
  \end{center}
\end{figure}

Varying the value of the parameter $b$, we obtain a prediction of the model as curves in the $(n_{\text{s}}, r)$-plane in Fig.~\ref{fig:ns_r}.  
The region of positive $b$ corresponds to the left of the large point (the Starobinsky model, $b=0$), and negative to the right.
This is because the positive $b$ makes the potential flatter ($\epsilon$ smaller) and more curved ($|\eta|$ larger), so both $n_{\text{s}}$ and $r$ are smaller.
The constraint on $n_{\text{s}}$ gives a constraint on $b$ as the inequalities~\eqref{b_bound50} and \eqref{b_bound60}.
The correction from the extra term does not drastically change the value of $r$ to improve the detection prospect of $r$.
However, the future prospect of precision of the spectral index $n_{\text{s}}$ by the 21 cm line and CMB observation will be $5\times 10^{-4}$~\cite{Kohri:2013mxa}, so it will become possible to distinguish the Starobinsky model ($b=0$) and our extension if $b$ is at least of order $10^{-5}$.

\begin{figure}[bthp]
  \begin{center}
    \includegraphics[clip,width=10.0cm]{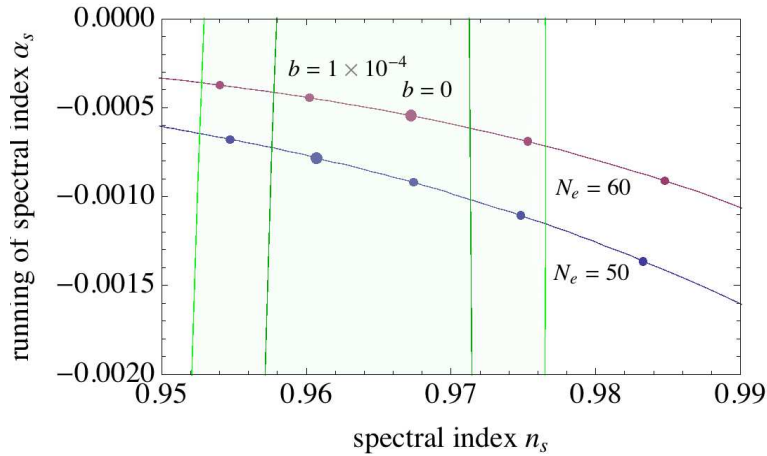}
    \caption{Prediction of the model in the ($n_{\text{s}}, \alpha_{\text{s}}$)-plane as we vary the parameter $b=b_3 b_1$. 
    The blue (bottom) line and red (top) line correspond to $N_{e}=50$ and $N_{e}=60$, respectively.
    The large dots show the prediction of the Starobinsky model ($b=0$), and the small dots correspond to the points of $b= n \times 10^{-4}$ with integer $n$.
    The green contours are the Planck TT, TE, EE $+$ low P constraints (traced from Fig. 4 in Ref.~\cite{Ade:2015lrj}).}
    \label{fig:ns_alphas}
  \end{center}
\end{figure}

The prediction of the model in the $(n_{\text{s}}, \alpha_{\text{s}})$-plane is shown in Fig.~\ref{fig:ns_alphas}.  
As can be seen from the figure and Eq.~\eqref{alphas}, the positive $b$ makes the absolute value of the running smaller.
The running in the model~\eqref{V} with the rising correction ($b<0$) was discussed in Ref.~\cite{Broy:2014xwa} in the context of power suppression in low multipoles.\footnote{
It is interesting to note that such steepening arises also in the (old-minimal) supergravity embedding of the Starobinsky model for some initial conditions~\cite{Kallosh:2014qta, Ellis:2014rxa, Hamaguchi:2014mza}.
}
The running $\alpha_{\text{s}}$ will be also measured by the 21 cm line and CMB observation at the precision of $3\times 10^{-4}$~\cite{Kohri:2013mxa}, and this has sensitivity on our deformation parameter $b$ of order $10^{-4}$.
This information combined with that of $n_{\text{s}}$ will give us a test of our model.

On the other hand, the running of running $\beta_{\text{s}}$ is of order $10^{-5}$ in our case, and it cannot be measured by the near future observations~\cite{Kohri:2013mxa}.
In other words, our predictions can be falsified by detection of the running of running $\beta_{\text{s}}$.

Let us discuss the uncertainties due to the reheating epoch in our predictions.
The uncertainty of the $N_e$ affects $n_{\text{s}}$ in particular, see Figs.~\ref{fig:ns_r} and \ref{fig:ns_alphas}.
We estimate it for the original Starobinsky model as $N_e = 56.0(1)$ where we have used the expression of $N_e$ in Ref.~\cite{Planck:2013jfk} and the inflaton decay rate in Ref.~\cite{Gorbunov:2010bn}, we have assumed the equation-of-state parameter during reheating as $w=0$ due to the coherent oscillation of the inflaton field, and the uncertainty estimate is only of order of magnitude. 
The analysis in Ref.~\cite{Gorbunov:2010bn} is based on the perturbative decay.
Nonperturbative production of the inflaton particles, in particular localized objects called oscillons or $I$ balls, (as well as light particles) is ineffective even when taking the metric preheating into account~\cite{Takeda:2014qma}.
The dominant source of the uncertainty of $N_e$ comes from that of the normalization of the CMB power spectrum.
The effect of the uncertainty on $n_s$ is roughly the same magnitude as that from higher-order corrections in terms of $N_e$, but the latter can be improved relatively easily (\textit{e.g.} by numerical calculation).
The higher-order corrections for the Starobinsky model are available in Ref.~\cite{Kaneda:2010ut}.

When introducing the deformation parameter $b$, the effect of changing the shape of the potential contributes to the uncertainty at most $\Delta N_e = \mathcal{O}(10^{-2})$.
The dominant source of uncertainty may come from change of the reheating temperature depending on the assumptions on inflaton-matter couplings.
If matter fields couple to gravity minimally after dimensional reduction to four dimensions, the inflaton-matter couplings are the usual Planck-suppressed couplings due to Weyl rescaling.
If there are interaction vertices suppressed by $\Lambda^{-1}$, the reheating temperature is expected to be scaled as $M_{\text{P}}/\Lambda$.
Since $\Lambda$ is related to $b_1$ as in Eq.~\eqref{first_2parm_fit}, $N_e$ increases by 1.9 to 2.7 as $|b_1|$ is varied from $10^{-4}$ to $10^{-2}$ (\textit{cf.}~footnote~\ref{ftn:tuning}). 
Combining these discussions with Figs.~\ref{fig:ns_r} and \ref{fig:ns_alphas}, we conclude that we can distinguish our deformation from the original Starobinsky model if $|b_1|$ is of order $10^{-4}$, but $|b_1|$ of order $10^{-5}$ seems difficult to distinguish taking the uncertainties into account.

If there is a nonminimal coupling $\xi h^2 R$ between a scalar (\textit{e.g.}, a Higgs boson) $h$ and Ricci scalar $R$, many quanta of $h$ can be produced by preheating with $|\xi|\gtrsim 5$~\cite{Tsujikawa:1999iv}.  
However, if we adopt the maximal production case to the Standard Model Higgs, the electroweak vacuum will become unstable, which should be avoided.
The value of $\xi$ is constrained to be at most of order $1$.
Precise estimation of the reheating temperature in this case is subject to the uncertainties of the top mass, the Higgs mass, and the strong coupling constant through the renormalization group running of the Higgs quartic coupling.
For example, if we take $\lambda = 5\times 10^{-3}$ at $\mu = \sqrt{\langle h^2 \rangle } = 10^{10}$~GeV (near the central line of the right panel of Fig.~1 in Ref.~\cite{Degrassi:2012ry}; $\mu$ is the renormalization scale), the preheating effect is subdominant compared to the perturbative decay as discussed above. On the other hand, with the same value of $\lambda$ at $\mu=\sqrt{\langle h^2 \rangle } = 10^{12}$~GeV (near the edge of the 3$\sigma$ band of the right panel of Fig.~1 in Ref.~\cite{Degrassi:2012ry}), we find $\Delta N_e \simeq 0.9$.
Thus, once we assume existence of the nonminimal coupling, the small change of the observables due to the modification parameter $b$ may or may not be buried in the effect of the nonminimal coupling $\xi$, depending on the precise values of the electroweak observables.

%%%%%%%%%%%%%%%%%%%%%%%%%%%%%%%%%%%%%%%%%
\section{Discussion} \label{sec:discussion}
In this paper, we proposed a new interpretation of the Starobinsky model as a low-energy effective theory of 
a higher-dimensional theory whose characteristic energy scale is denoted by $\Lambda$.
Compactification of extra dimensions naturally introduces a large overall factor.
With the tuning of $|b_1|\lesssim  2\times 10^{-4}$ (in addition to the one for the cosmological constant), we obtain the Starobinsky model augmented with higher-order terms suppressed enough to be consistent with the Planck 2015 results.
Compared to taking a large parameter only in front of the $R^2$ term, taking a small parameter is regarded as less unnatural in the sense that it may happen by accidental cancelation of several contributions.
We have also argued that the original version suffers from tuning of infinite parameters and infinite precision, whereas finite tuning of one parameter is enough in our framework.
If the deformation is indeed of order $10^{-4}$, the model can be distinguished from the original Starobinsky model ($b=0$) by future observations of CMB and the 21 cm line.

Predictions on inflationary observables studied in the previous section are consequences of the action~\eqref{CompactifiedAction}, but do not crucially depend on the underlying assumption~\eqref{DdimAction}.
Here, we briefly discuss another possibility to obtain the advocated action~\eqref{CompactifiedAction}.
One of the reasons for the unnatural expansion of the Starobinsky model action may reside in the fact that we regard the Einstein term as the fundamental term, and the other terms are ``secondary'' 
  in the sense that they originate from quantum corrections.
In contrast, we can take a view that the fundamental or main term is the second term $R^2$ rather than $R$ (see \textit{e.g.}, Refs.~\cite{Gorbunov:2013dqa, Salvio:2014soa, Kannike:2015apa, Bamba:2014mua, Kounnas:2014gda, Rinaldi:2014gha, Einhorn:2014gfa} in this kind of direction).
The pure $R^2$ theory, $S= c \int d^4 x \sqrt{-g}  R^2$, does not have a dimensionful constant, and it is scale invariant.\footnote{
Aspects of quadratic gravity were recently revisited in Ref.~\cite{Alvarez-Gaume:2015rwa}, and supergravity embedding of the pure $R^2$ theory was studied in Ref.~\cite{Ferrara:2015ela}. 
See also the inflation scenario based on broken scale invariance in Ref.~\cite{Csaki:2014bua}.
}
Inflation in this theory is in the pure de Sitter universe, and it eternally inflates.
Note that the coefficient $c$ of the action cannot be absorbed into $R$ by scale transformation simply because the action is scale invariant, and it is legitimate to take the coefficient as a huge or minuscule number.  The former eventually corresponds to the Starobinsky model.
If the scale symmetry is spontaneously broken, perhaps after coupling to matter sector, then a scale $\Lambda$ is generated.
This will lead to the form of the action~\eqref{CompactifiedAction}.
The fact that $|b_1|$ should be suppressed is unchanged, and we have the same predictions \eqref{ns}, \eqref{r}, \eqref{alphas}, and \eqref{betas}.
In this context, smallness of $|b_1|$ corresponds to soft breaking of the scale symmetry, and hence is technically natural.
This is same in our scenario:  After compactification, the approximate scale symmetry appears with a small symmetry breaking parameter.
With technical naturalness combined with the explanation of the huge coefficient of $R^2$ in terms of extra dimensions, the form of the Starobinsky model is naturally understood.

It would also be useful to discuss another possible explanation for
the tuning of $b_1$ (the cosmological constant also).
From the effective theory point of view,
the first few terms in the low energy expansion
have very small values in our model.
Indeed,
such a situation sometimes occurs
in the effective action of the order parameter near the phase transition point (e.g. the Lifshitz point of the Nambu--Jona-Lasinio model~\cite{Nambu:1961tp, Nambu:1961fr})~\cite{Nickel:2009ke, Nickel:2009wj}.
Based on this analogy,
let us think of the metric as some ``order parameter"
and explore the ``phase structure" of $f(R)$ gravity,
using the action $S= \int d^4 x \sqrt{-g} c (-d_0-d_1 R+R^2)$ with $c>0$.
Under the standard procedure, we rewrite this as a theory of a canonical scalar field in the Einstein frame.
The potential is bounded below as long as $d_1^2-4d_0>0$.
The shape of the scalar potential dramatically changes depending on the sign of $d_1$: 
(i) For $d_1=0$, the potential is constant, and its sign (dS, AdS, or Minkowski) is determined by the value of $d_0$.
(ii) For $d_1>0$, just as in the Starobinsky model, the potential has a minimum as well as a flat region.
Inflation is therefore realized.
(iii) For $d_1<0$, the minimum of the potential is at $\phi=\infty$ (runaway potential).
The inflaton slow-rolls toward $\phi=\infty$ and the eternal inflation in the approximate dS is realized.
It would be interesting if we could interpret inflation as a consequence of the phase transition from phase (ii) through (i) to (iii).

Finally,
let us estimate the characteristic scale $\Lambda$ in the higher dimensional theory.
Using the upper bound on $|b_1|$,
its lower bound is given by
\begin{align}
\Lambda=m\sqrt{\frac{6}{|b_1|}}\gtrsim 5 \times 10^{15} \text{GeV}.
\end{align}
This scale is close to the Grand Unified Theory scale, so it is tempting to relate the higher dimensional theory with Grand Unification.
Moreover, the Planck scale $M_{\text{P}}$ is not the fundamental scale in our viewpoint, and
quantum gravity effects may appear at this scale.
In the context of superstring theory, we can identify the scale $\Lambda$ as the string scale $m_{\text{s}}$ (see the discussion after Eq.~\eqref{CompactifiedAction}).
Furthermore, modification of gravity at a length scale larger than the Planck length (or in other words, low cut-off theory) is favored from a variety of perspectives including a solution to the cosmological moduli problem~\cite{Nakayama:2011zy} and a way to preserve global symmetries to a good precision~\cite{Kallosh:1995hi}, which is crucial, \textit{e.g.}, for the axion solution to the strong CP problem~\cite{Kim:1986ax}.
These discussions imply that we may see the footprints of quantum gravity or string theory in the sky, which might simultaneously give us some hints on cosmological and particle physics problems.

\section*{Acknowledgement}
We thank the organizers of the ``Workshop to propose ways of testing new physics'' and the staff in the Okinawa Institute of Science and Technology Graduate University (OIST) where the workshop took place and we initiated this project.
The workshop was supported by the JSPS Grants-in-Aid for Scientific Research No.~23740192, No.~25105011, and No.~23740165.
This work was supported by JSPS Grants-in-Aid for Scientific Research No.~25400249 (T.A.), No.~26105508 (T.A.), No.~15H01031 (T.A.), No.~26105520 (K.K.), No.~26247042 (K.K.), No.~15H05889 (K.K.), No.~JP16H00877 (K.K.) and No.~26$\cdot$10619 (T.T.).
The work of S.I.~and K.K.~was also supported by the Center for the Promotion of Integrated Science (CPIS) of
Sokendai (1HB5804100).
T.N.~was supported by Special Postdoctoral Researchers Program at RIKEN and the RIKEN iTHES Project.
T.T.~was also supported by a Grant-in-Aid for JSPS Fellows.
This work is also supported in part by National Research Foundation of Korea (NRF) Research Grant NRF-2015R1A2A1A05001869 (T.T.).

\section*{Funding}

\noindent{Open Access funding: SCOAP$^3$.}

\noindent{Japan Society for the Promotion of Science.}

\noindent{RIKEN.}

\noindent{National Research Foundation of Korea.}


%\bibliographystyle{ptephy}
%\bibliography{ref}
%
% once the .bbl file has been generated then place the text in your article.

\end{document}